\documentclass[aps,prd,twocolumn,a4paper]{revtex4-1}
\usepackage{ulem}
\usepackage{color}
\usepackage{graphicx}
\usepackage{epsfig}
\newcommand{\be}{\begin{equation}}
\newcommand{\ee}{\end{equation}}
\newcommand{\ba}{\begin{eqnarray}}
\newcommand{\ea}{\end{eqnarray}}
\newcommand{\nn}{\nonumber\\}
\begin{document}
\begin{flushright}
{Report No.:TIFR/TH/11-50}
\end{flushright}
\title{Transport properties of anisotropically expanding quark-gluon plasma within a quasi-particle model} 
\author{Vinod Chandra$^{a,b}$}
\email{vinod.chandra@fi.infn.it\\
$^b$ is the current affiliation of the author
}
\affiliation{$^a$ Department of Theoretical Physics, Tata Institute of Fundamental Research,
Homi Bhabha Road, Mumbai-400005, India.}
\affiliation{$^{b}$ Istituto Nazionale di Fisica Nucleare (INFN) Sezione di Firenze, Italy}

\date{\today}
\begin{abstract}
The bulk and shear viscosities ($\eta$ and $\zeta$) have been studied 
for quark-gluon-plasma produced in relativistic heavy ion collisions within semi-classical transport 
theory, in a recently proposed quasi-particle model of (2+1)-flavor lattice QCD equation of state. 
These transport parameters have been found to be highly sensitive to the interactions present in hot QCD.
Contributions to the transport coefficients from both the gluonic sector and the matter sector have been 
investigated. The matter sector is found to be significantly dominating over the gluonic sector, in both the cases of 
$\eta$ and $\zeta$. The temperature dependences of the quantities, $\zeta/{\mathcal S}$, and   $\zeta/\eta$ indicate a sharply 
rising trend for the $\zeta$, closer to the QCD transition temperature. Both $\eta$, and $\zeta$ are shown to be equally significant for the temperatures that are accessible in the relativistic heavy ion collision experiments, and hence play crucial role while investigating the properties of the quark-gluon plasma.
  
\vspace{2mm}

\vspace{2mm}
\end{abstract}
\maketitle

\section{Introduction}
The study of transport coefficients for hot QCD matter is 
an area of intense research since the discovery of a fluid like 
picture of quark-gluon-plasma (QGP) in the relativistic heavy ion collider (RHIC) 
at BNL~\cite{expt}. The discovery of the QGP is attributed to the fact that 
at extreme energy-density and temperature, ordinary nuclear
matter goes through a transition to the QGP phase as predicted by the finite temperature Quantum-Chromodynamics (QCD)
(this transition is shown to be a crossover~\cite{cross} at the vanishing baryon density).

To describe a fluid, shear and bulk viscosities ($\eta$ and $\zeta$ respectively) 
are very important physical quantities 
that characterize dissipative processes during its hydrodynamic evolution. 
The former describes the entropy production due to the transformation of the shape of hydrodynamic
system at a constant volume, and the latter describes the entropy production at the constant rate 
of change of the volume of the system (hot fireball at the RHIC). 
Moreover, $\eta/{\mathcal S}$, and $\zeta/{\mathcal S}$ serve as the inputs 
while studying the hydrodynamic evolution of the fluid~\cite{shrhydro,bulkhydro}. One can 
also couple hydrodynamics with the Boltzmann descriptions at the later stages after the collisions 
of heavy ions at the RHIC, by maintaining the continuity of the entire stress energy tensor and currents.
The process could be translated in terms of the viscous modifications to the thermal distributions functions 
of particles. This leads to a smooth transition from the hydrodynamic regime where the mean free paths are
short to a region where hydrodynamics is inapplicable and Boltzmann treatments seems to be justified~\cite{prat}.
Therefore, this sets a way to study the impact of transport coefficients of the QGP 
in various processes at the RHIC, and the ongoing heavy ion experiments at Large Hadron Collider (LHC), CERN ({\it e.g.} dilepton production, quarkonia physics {\it etc.}). Regarding viscous corrections to dilepton production rate at the RHIC, we refer the reader to~
\cite{srikanth}. The determinations of $\eta$ and $\zeta$ have to be done separately from a microscopic theory;
either from a transport equation~\cite{landau} with an appropriate force, collision, and source terms or  equivalently from the field theoretic approach by employing
the Green-Kubo formulas~\cite{green-kubo} (long wavelength behavior of the correlations among 
various components of the stress-energy tensor). 

The QGP is strongly interacting at the RHIC~\cite{expt}, as inferred from the flow measurements,
and strong jet quenching observed there. This observation is found to be consistent  with the 
lattice simulations of the hot QCD equation of state (EOS)~\cite{leos_lat,leos1_lat}, which predict 
a strongly interacting behavior even at temperatures which are of the order of a few $T_c$ (the QCD transition temperature). 
The flow measurements suggest a very tiny value for the ratio of $\eta$ to the entropy density,
${\mathcal S}$ ($\eta/{\mathcal S}$) for the QGP, and the near perfect fluid picture~\cite{shrvis,bmuller,chandra_eta1,chandra_eta2}
(except near the QCD transition temperature where $\zeta/{\mathcal S}$ is equally significant as $\eta/{\mathcal S}$~\cite{khaz,mayer,buch,chandra_bulk}).

Preliminary studies at the LHC~\cite{alice,alice1,hirano} reconfirm above mentioned observations regarding 
the QGP. In heavy-ion collisions at the LHC, in addition to the elliptic flow obtained at the RHIC,
there are other interesting flow patterns, {\it viz.}, the dipolar, and the triangular flow, which are sensitive to the initial collision geometry~\cite{flow_lhc}. There have been recent interesting studies to understand them at LHC~\cite{bhalerao,alice}. 
A more precise measurement of various flows and jet quenching at LHC is awaited.
On the other hand, $\zeta$ has achieved considerable attention in the context of the QGP 
after the interesting reports on its rising value close to the QCD transition temperature~\cite{khaz}.
Subsequently, the possible impact of the large bulk viscosity of the QGP at the RHIC have been studied
by several authors; Song and Heinz~\cite{heinz} have studied the interplay of 
shear and bulk viscosities in the context of collective flow in 
heavy ion collisions. Their study revealed that one can not simply 
ignore the bulk viscosity while modeling the QGP.
In this context, there are other interesting studies in the literature~\cite{den,raj1,hirano1,raj,efaaf,pion,fries}.
The role of bulk viscosity in freeze out phenomenon 
has been offered in~\cite{torri,hirano}. Effects of bulk viscosity in the hadronic phase,  and in the hadron emission
have been studied in~\cite{boz}. Interestingly, in the recent investigations, these transport coefficients are found 
to be very sensitive to the interactions~\cite{chandra_eta1,chandra_eta2}, and 
the nature of the phase transition in QCD~\cite{moore}. 
Another crucial aspect of $\zeta$ is its influence on the 
domain of applicability of hydrodynamics at the RHIC, {\it viz.} the phenomenon 
cavitation. This phenomenon has been addressed in detail in the 
context of diverging value of $\zeta$ near the QCD transition temperature 
in~\cite{rajgopal,srikanth1}. Thus, the determinations of $\eta$ and $\zeta$ for 
the QGP have multi-facet dimensions, and significant impact on the variety of physical phenomena 
at the RHIC and the LHC. 
Subsequently, the cavitation in a particular string theory model (N=2* SU(N) theory
which is non-conformal, and mass deformation of N=4, SU(4) Yang-Mills) has been investigated by Klimek, Leblond, and Sinha
in~\cite{sinha}. They have observed the absence of cavitation before phase transition is reached, by investigating 
the flow equations in (1+1)- dimensional boost invariant set up, which is in contrast to the 
finding of~\cite{rajgopal} for hot QCD. They further argued that such a behavior is mainly due to to
smaller value of 
$\zeta$, and sharp rise of the relaxation time for such theories near the transition point, and perhaps the 
quantum corrections to $\eta$, and $\zeta$~\cite{sinha_qnt}.  These studies might play a crucial role in
understanding the behavior of strongly coupled QGP (sQGP) in the RHIC and the LHC.

The determinations of $\eta$ and $\zeta$ have been performed adopting the 
view-point based on the inference drawn from the 
experimental results, and the lattice QCD (the best known non-perturbative technique
to address the QGP). Lattice QCD has indeed been very successful to study the 
QGP thermodynamics. However, the the computation of the transport coefficients in lattice QCD 
is a very non-trivial exercise, due to several uncertainties and 
inadequacy in their determination. Despite that there are a few first results computed from lattice QCD for bulk and 
shear viscosities~\cite{meyer,tmr,meyer1,nakamura} which have observed a small value of $\eta/{\mathcal S}$,  and 
large  $\zeta/{\mathcal S}$ at the RHIC. A very recent interesting 
analysis~\cite{gupta} suggests that it is possible to compare the 
direct lattice results with the experiments at the RHIC. From such a comparison, the QCD transition temperature came 
out to be around $175 MeV$. More refined lattice studies on $\eta$ and $\zeta$ are awaited in the near future.

The work presented in this paper is an attempt to achieve, (i)
temperature dependence of $\eta$ and $\zeta$ (The gluonic as well as the matter sector 
contributions to these transport parameters have been obtained by combing a transport equation with a
recently proposed quasi-particle model~\cite{chandra1,chandra2,chandra_quasi} of $(2+1)$-flavor lattice QCD EOS.
Noteworthy  point is that the matter sector has largely been
 ignored in the literature in this context), (ii) to understand the small $\eta/{\mathcal S}$, and  large $\zeta/{\mathcal S}$ for the QGP 
for the temperatures closer to $T_c$. 
More precisely, inputs has been taken from the computations of $\eta$ and $\zeta$ 
in  quasi-particle models~\cite{chandra_eta1,chandra_eta2,kapu,sakai,chandra_bulk}, 
and combine the understanding with a transport 
theory determination of them in the presence of chromo-Weibel instabilities~\cite{bmuller,chromw,bmuller1}.
The present work is the extension of our recent work on $\eta$~\cite{chandra_eta1,chandra_eta2}, 
and $\zeta$~\cite{chandra_bulk} for the gluonic sector, to the (2+1)-flavor QCD.

The paper is organized as follows. In Sec. II, we present the formalism to compute the 
$\eta$ and $\zeta$. The quasi-particle model and transport equation have also been discussed 
in brief in the same section. In Sec. III, we have presented the results on the 
temperature dependence of $\eta$ and $\zeta$ in $(2+1)$-flavor lattice QCD, and relevant physics.
In Sec. IV, we have presented conclusions and future prospects of the present work.

\section{Determination of transport coefficients}
There may be a variety of physical phenomena that lead to the viscous effects in the QGP (or in general any interacting system)
~\cite{prat}. Among them, our particular focus is on the viscous effects which
 get contributions from the the classical chromo-fields. 

The idea adopted here is based on the mechanism earlier proposed in
~\cite{bmuller,bmuller1,majumdar} to explain the small viscosity of a weakly coupled, but expanding QGP.
The mechanism in the context of the QGP is solely based 
on the particle transport processes in the turbulent plasmas~\cite{dupree} that are 
characterized by strongly excited random field modes in the certain regimes of instability.
They coherently scatter the charged particles, and thus reduce the rate of momentum
transport. This eventually lead to the suppression of the transport coefficients in plasmas.
This phenomenon has been studied both in electro-magnetic (EM) plasmas~\cite{niu}, and
in non-abelian plasmas (QCD plasma) by Asakawa, Bass and M\"{u}ller~\cite{bmuller,bmuller1}, 
and further employed for the realistic QGP EOS in~\cite{chandra_eta1,chandra_eta2}.

The condition for the spontaneous formation of turbulent fields can be 
achieved in  EM plasmas with an anisotropic momentum 
distribution~\cite{weibel} of charged particles, and in the QGP with anisotropic distribution of thermal 
partons~\cite{sma}. In the context of pure SU(3) gauge theory, 
this mechanism turn out to be successful to explain small shear viscosity 
of the QGP and larger bulk viscosity for the temperatures accessible at the RHIC and the LHC~\cite{chandra_eta2,chandra_bulk}.
Here, extension has been desired to the case of realistic EOS for the QGP by incorporating the effects 
from the matter sector (quark-antiquarks).

It will be seen later that the analysis leads to an interesting observation 
regarding the relative contribution of the gluonic and the matter sectors to the transport 
parameters. Before, we present a brief 
description of the quasi-particle understanding of $(2+1)$-flavor lattice QCD that furnishes an appropriate 
modeling of equilibrium state.

\subsection{The quasi-particle description of hot QCD}
Quasi-particle description of the hot QCD medium effects, is not a new concept.
There have been several attempts so far to understand the hot QCD medium effects in terms of 
non-interacting/weakly interacting quasi-partons, {\it viz.}, 
effective thermal mass models~\cite{pesh,pesh1},  effective mass model 
with temperature dependent bag parameter to cure the problem 
of thermodynamic inconsistency~\cite{pesh1}, effective quasi-particles 
with gluon condensate~\cite{glucond}, Polyakov loop models~\cite{polya}
(Polyakov loop acts as effective fugacity), and the quasi-partons with effective fugacities
~\cite{chandra1,chandra2,chandra_quasi}. The last one that will 
be employed here, shown to be fundamentally distinct from all other mentioned models,
and in the spirit of Landau's theory of Fermi liquids. Moreover, the model 
has been highly successful in interpreting the lattice QCD thermodynamics, and
bulk and transport properties of hot QCD matter and the QGP in relativistic heavy ion collisions.

In our quasi-particle description for (2+1)-flavor lattice QCD~\cite{chandra_quasi}, we start with the ansatz that
the Lattice QCD EOS can be interpreted in terms of non-interacting
quasi-partons having effective fugacities which encode all the interaction effects. 
We denote them as gluon effective fugacity, $z_g$ and the quark-antiquark fugacity, $z_q$.
In this approach, the hot QCD 
medium is divided in to two sectors, {\it viz.}, the effective gluonic sector, and 
the matter sector (light quark sector, and strange quark sector). The former refers to the contribution of 
gluonic action to the pressure which also involves contributions 
from the internal fermion lines. On the other hand, latter involve interactions among quark, anti-quarks, as well as  
their interactions with gluons.  The ansatz can be translated to the form of the equilibrium distribution functions, 
$ f_{eq}\equiv \lbrace f_{g}, f_{q}, f_{s} \rbrace$ (this notation will be useful later while 
writing the transport equation in both the sector in compact notations)  as follows,

\ba
\label{eq1}
f_{g} &=& \frac{z_g\exp(-\beta p)}{\bigg(1-z_g\exp(-\beta p)\bigg)},\nn
f_{q} &=& \frac{z_q\exp(-\beta p)}{\bigg(1+z_q\exp(-\beta p)\bigg)},\nn
f_{s} &=& \frac{z_q\exp(-\beta \sqrt{p^2+m^2})}{\bigg(1+z_q\exp(-\beta \sqrt{p^2+m^2})\bigg)},
\ea
where $m$ denotes the mass of the strange quark, which we choose to be $0.1 GeV$. The parameter, $\beta=T^{-1}$
denotes inverse of the temperature. Here, we are working in the units where Boltzmann constant, $K_B=1$, $c=1$, and $h/2\pi=1$.
The notation $p$ is nothing but, $p\equiv \vert \vec{p} \vert$.

We use the notation $\nu_g=2(N_c^2-1)$ for gluonic degrees of freedom , $\nu_{q}=2\times 2\times N_c\times 2$
for light quarks, $\nu_s=2\times 2 \times N_c \times 1$ for the strange quark for $SU(N_c)$. 
Here, we are dealing with $SU(3)$, so  $N_c=3$. Since the model is valid in the deconfined phase of QCD (beyond $T_c$), therefore, the mass contributions of the light quarks can be neglected as compared to 
the temperature. Therefore, in our model, we only consider the mass for the strange quarks.  

The effective fugacity is not merely a temperature dependent parameter which encodes the 
hot QCD medium effects. It is very interesting and physically significant.
The physical significance reflects in the modified dispersion relation both in the 
gluonic and matter sector. In this description, the effective fugacities modify the single quasi-parton energy
as follows,
\ba
\label{eq2}
\omega_g&=&p+T^2\partial_T ln(z_g)\nn
\omega_q&=&p+T^2\partial_T ln(z_q)\nn
\omega_s&=&\sqrt{p^2+m^2}+T^2\partial_T ln(z_q).
\ea

These dispersion relations can be explicated as follows. The single quasi-parton energy not only depends upon its momentum
but also gets contribution from the collective excitations of the quasi-partons.
The second term is like the gap in the energy-spectrum due to the presence of 
quasi-particle excitations. This makes the model more in the spirit of the Landau's theory of Fermi -liquids.
For a detailed discussion on
the interpretation and physical significance of $z_g$, and $z_q$, we refer the reader 
to our recent work~\cite{chandra_quasi}. Henceforth, we shall use gluonic sector in the place of effective 
gluonic sector for the sake of ease. We shall now proceed to the determination $\eta$ and $\zeta$ 
in presence of chromo-Weibel instabilities.

\subsection{Chromo-Weibel instability and the anomalous transport}
The determinations of $\eta$ and $\zeta$ have been done in a  multi-fold way.
Firstly, we need an appropriate modeling of distribution functions for the equilibrium state.
Secondly, we need to set up an appropriate transport equation to determine the 
form of the perturbations to the distribution functions. These two steps eventually determine these
transport coefficients. For the former step, we employ the quasi-particle model for 
the (2+1)-flavor lattice QCD EOS discussed earlier.

Both $\eta$ and $\zeta$ have two contributions, same as in the case the shear viscosity
 in~\cite{bmuller}, (i)
from the Vlasov term which captures the long range component
of the interactions, and (ii) the collision term, which
models the short range component of the interaction. 
Here, we shall only concentrate on the former case. 
The determinations of shear and bulk viscosities 
from an appropriate collision term will be a matter 
of future investigations. Importantly, the analysis adopted here is 
based on weak coupling limit in QCD, therefore, the results are shown beyond 
$1.2 T_c$ assuming the validity of weak coupling results for the QGP there.
Note that the interplay for anomalous and collisional components 
of $\eta$ has been discussed in~\cite{bmuller,chandra_eta1,chandra_eta2}, and 
in the case of $\zeta$ for the pure gauge theory, a discussion has been presented regarding 
the interplay of the collisional~\cite{wang_bulk,wang_shear,arnold},
and anomalous components in~\cite{chandra_bulk}. 
It seems that at the conceptual level all the observation in ~\cite{chandra_bulk}
regarding the interplay will remain valid here. Since, we do not have results for the matter sector 
therefore, we shall not offer a quantitative discussions on such an interplay here. 
There have been computations of transport parameters in the case of pure gauge theory 
based on the effective mass models within the relaxation time approximation~\cite{transport_quasi}. 
The approach adopted, and the physical set up is entirely distinct 
in the present case. It is to be noted that the gluonic component in all 
the quantities is denoted by sub/superscript $g$, similarly for light-quark components by $q$, and strange quark component by $s$
respectively.

\subsection{Determination of $\zeta$ and $\eta$}
Let us first briefly outline the standard procedure of determining transport coefficients in transport theory~\cite{landau,bmuller}. The bulk and shear viscosities, $\zeta$ and $\eta$ of the QGP in terms of 
equilibrium parton distribution functions are obtained by comparing the kinetic theory definition   of the stress tensor with the fluid dynamic definition of the viscous stress tensor.

In kinetic theory, the stress tensor is defined as
\begin{equation}
\label{eq3}
 T^{\mu\nu}=\sum \int \frac{d^3\vec{p}}{(2\pi)^3 \omega} p^\mu p^\nu  f(\vec{p},\vec{r}),
\end{equation}
where the sum is over all species (in the present case, gluons, light-quarks and strange quarks)
including the internal degrees of freedom which is implicit in Eq. (\ref{eq3}).
The quantities $\omega\equiv \lbrace \omega_{g},\omega_q, \omega_s\rbrace$
combindly denote the quasi-particle dispersions, and $f(\vec{p},\vec{r})$  is the combined
notation for the quasi-particle distribution functions.

This form of $T^{\mu\nu}$ does not capture the medium modifications encoded in the non-trivial 
dispersion relations, $\omega$ and hence does not implement the thermodynamic consistency condition correctly.
This is very crucial in its own merit, and also needed to relate to the 
hydrodynamic definition of $T^{\mu\nu}$. In the present case, to obtain the correct expression of the energy density, one needs
to modify the 4-momenta of the quasi-particles, which is not allowed in the model in view of the particular mathematical 
structure of the equilibrium distribution functions in Eq. (\ref{eq1}). 
To cure the problem, the definition of $T^{\mu\nu}$ need to be modified such that 
$u_\mu u_\nu T^{\mu\nu}=\epsilon$ (true energy density). This can be achieved by 
the revised definition of $T^{\mu\nu}$ in case of our quasi-particle model with 
effective fugacities,

\begin{eqnarray}
\label{eq3a}
T^{\mu\nu}&=&\sum \bigg\lbrace \int \frac{d^3\vec{p}}{(2\pi)^3 \omega} p^\mu p^\nu f(\vec{p},\vec{r})\nonumber\\
          &&+ \int \frac{d^3\vec{p}}{(2\pi)^3 p \omega} (\omega-E_p) p^\mu p^\nu f_0(\vec{p},\vec{r})\nonumber\\
          &&+ \int \frac{d^3\vec{p}}{(2\pi)^3} (\omega-E_p) u^\mu u^\nu f_0(\vec{p},\vec{r})\bigg\rbrace,
\end{eqnarray}
where $E_p$ denote the dispersions without medium modifications, $E_p=p$ for gluons, and light quarks, and 
$E_p=\sqrt{p^2+m^2}$ for the s-quarks, and antiquarks respectively. Therefore, one can clearly realize the 
presence of the factors, $T^2 \frac{d ln(z_g)}{d T}$, and $T^2 \frac{d ln(z_q)}{d T}$ in the
expression for $T^{\mu\nu}$. The second term in the right-hand side of Eq. (\ref{eq3a}) ensures the 
correct expression for the pressure, and the third term ensures the correct expression for the 
energy density, and hence the definition of $T^{\mu\nu}$ incorporates the thermodynamic consistency condition
correctly. This issue is realized in a similar way in the effective mass quasi-particle models in~\cite{dush2},
accordingly the modified definition of $T^{\mu\nu}$ is employed that contains temperature derivative of 
effective mass.

On the other hand, in hydrodynamics the expression for the  viscous stress tensor up to first order 
in the gradient expansion is given by,
\begin{equation}
\label{eq4}
T^{\mu\nu}=(\epsilon+P)  u^\mu u^\nu -P g^{\mu\nu} -\Pi \Delta^{\mu\nu}+\pi^{\mu\nu},
\end{equation}
where, $u^\mu$ is the fluid 4-velocity, $g^{\mu\nu}$ is the metric tensor,
$\Delta^{\mu\nu}=g^{\mu\nu}-u^\mu u^\nu$ is the orthogonal projector, $\Pi$ is the bulk 
part of the stress tensor, and $\pi^{\mu\nu}$ is the shear stress. Here, $\epsilon$ is the energy-density and $P$ is the pressure
of the fluid.

In the first order (Navier-Stokes) approximation, the 
viscous (dissipative) parts of the stress energy tensor in Eq.(\ref{eq4}),
can be obtained in local rest frame of the fluid (LRF) as,
\begin{eqnarray}
\pi_{ij}&=&-2\eta (\nabla u)_{ij}\nonumber\\
(\nabla u)_{ij}&=&\frac{\partial_i u_j+\partial_j u_i}{2}-\frac{1}{3} \delta_{ij} \partial_i u^j,\nonumber\\
\Pi&=&-\zeta \nabla\cdot \vec{u} \equiv \partial_k u^k,
\end{eqnarray}
where $(\nabla u)_{i k}$ is the traceless, symmetrized
velocity gradient, and $\nabla\cdot\vec{u}$ is the divergence of the fluid velocity field,
$\eta$ and $\zeta$ combindly denote the ($\eta_g,\eta_q, \eta_s$), and
($\zeta_g,\zeta_q, \zeta_s$) (later we shall write them explicitly).
In the LRF, ( $u^\mu=(1,0,0,0)$),
$f_0\equiv \lbrace f_g, f_q, f_s\rbrace$.

Next, to determine $\zeta$ an $\eta$, one writes the parton distribution functions as
\begin{equation}
\label{eq5}
f(\vec{p},\vec{r})=\frac{1}{z_{g/q}^{-1}\exp(\beta u^\mu p_\mu+f_1(\vec{p},\vec{r}))\mp 1}.
\end{equation}

Assuming that  $f_1(\vec{p},\vec{r})$ is a small perturbation to the equilibrium distribution, we expand $f(\vec{p},\vec{r})$ and keeping only the linear order term in $f_1$, one obtains,

\begin{eqnarray}
\label{eq6}
 f(\vec{p},\vec{r})&=&f_0(p)+\delta f(\vec{p},\vec{r})\nonumber\\
&=&f_0(p)\bigg(1+f_1(\vec {p},\vec{r})(1\pm f_0(p)\bigg),
\end{eqnarray}

where $f_0\equiv \lbrace f_g, f_q, f_s\rbrace$, and similarly $f_1\equiv \lbrace f^g_1,f^q_1,f^s_1 \rbrace$ 
in the LRF, and $p\equiv \vert \vec{p}\vert$ throughout the computations.
The {\it plus} sign in the bracket is for gluons, and {\it minus} sign is for fermions (q and s). Next, we shall consider these 
quantities explicitly in the gluonic and the matter sectors. As discussed in \cite{bmuller,chandra_eta2}, $\zeta$ and $\eta$ are determined by taking the following form of the perturbation $f_1$,

\begin{eqnarray}
\label{eq7}
 f^{g}_1(\vec{p},\vec{r})&=&-\frac{1}{\omega_{g} T^2} p_i p_j \bigg(\Delta_{1g}(p)
 (\nabla u)_{i j}+\Delta_{2g}(\vec{p}) (\nabla\cdot \vec{u})\delta_{ij}\bigg)\nonumber\\
 f^{q}_1(\vec{p},\vec{r})&=&-\frac{1}{\omega_{q} T^2} p_i p_j
 \bigg(\Delta_{1q}(p)(\nabla u)_{i j}+\Delta_{2q}(\vec{p})(\nabla\cdot \vec{u})\delta_{ij}\bigg)\nonumber\\
f^{s}_1(\vec{p},\vec{r})&=&-\frac{1}{\omega_{s} T^2} p_i p_j \bigg(\Delta_{1s}(p)(\nabla u)_{i j}+
\Delta_{2s}(\vec{p})(\nabla\cdot \vec{u})\delta_{ij}\bigg). \nonumber\\
\end{eqnarray}

Here, dimensionless functions $\Delta_{1g,1q,1s}(p), \Delta_{2g,2q,2s}(\vec{p})$  measure the deviation from the 
equilibrium configuration. $\Delta_1(p)$, $\Delta_2(\vec{p})$,  lead to $\eta$ and $\zeta$ respectively.
Note that  $\Delta_{1g,1q,1s}(p)$ is a isotropic function of the momentum in contrast
to $\Delta_{2g,2q,2s}(\vec{p})$, which is an anisotropic in momentum $\vec{p}$.  This is specifically associated
with the structure of the Vlasov operator in the present case. In this case, we seek a solution of the effective transport equation for the bulk viscosity that satisfies the LL condition, $u_\mu \delta T^{\mu\nu}=0$. To ensure that we have followed the description of Chakraborty, and  Kapusta~\cite{kapu} which has been discussed at the end of this section as Sec. IIE.

Since $\zeta$ and $\eta$ are  Lorentz scalars; they may be evaluated conveniently in the LRF
(in the LRF $f_0\equiv f_{eq}$). Considering the a boost invariant longitudinal flow, 
$\nabla\cdot \vec{u}=\frac{1}{\tau}$ and, $(\nabla u)_{i j} = \frac{1}{3\tau} diag (-1, -1,2)$ in the LRF.
In this case, the perturbations, $f_1(p)$ take the  form,

\begin{eqnarray}
\label{eq8}
f^{g}_1(\vec{p})&=&-\frac{\Delta_{1g}(p)}{\omega_{g} T^2\tau}\bigg(p_z^2-\frac{p^2}{3}\bigg)-\frac{\Delta_{2g}(\vec{p})}{\omega_g T^2\tau} p^2 \nonumber\\
f^{q}_{1}(\vec{p})&=&-\frac{\Delta_{1q}(p)}{\omega_{q} T^2\tau}\bigg(p_z^2-\frac{p^2}{3}\bigg)-\frac{\Delta_{2q}(\vec{p})}{\omega_{q} T^2\tau} p^2,\nonumber\\
f^{s}_{1}(\vec{p})&=&-\frac{\Delta_{1s}(p)}{\omega_{s} T^2\tau}\bigg(p_z^2-\frac{p^2}{3}\bigg)-\frac{\Delta_{2s}(\vec{p})}{\omega_{s} T^2\tau} p^2.
\end{eqnarray}

where $\tau$ is the proper time($\tau=\sqrt{t^2-z^2}$). 
The shear viscosities  are obtained in terms of entirely unknown functions, $\Delta_{1g,1q,1s}(p)$  as,

\begin{eqnarray}
\label{eq9}
\eta_g&=&\frac{\nu_g}{15 T^2}\int \frac{d^3\vec{p}}{8\pi^3} \frac{p^4}{\omega_g^2} \Delta_{1g}(p)f_{g}(1+f_{g})\nonumber\\
\eta_q&=&\frac{\nu_q}{15 T^2}\int \frac{d^3\vec{p}}{8\pi^3} \frac{p^4}{\omega_q^2} \Delta_{1q}(p)f_{q}(1-f_{q})\nonumber\\
\eta_s&=&\frac{\nu_s}{15 T^2}\int \frac{d^3\vec{p}}{8\pi^3} \frac{p^4}{\omega_s^2} \Delta_{1s}(p)f_{s}(1-f_{s})
\end{eqnarray}

The bulk viscosities are obtained in terms of the unknown functions, $\Delta_{2g,2q,2s}(\vec{p})$,

\begin{eqnarray}
\label{eq10}
\zeta_g&=&\frac{\nu_g}{3 T^2} \int \frac{d^3 \vec{p}}{8\pi^3} \frac{p^2}{\omega_g^2} (p^2-3 c^2_s \omega_g^2)\Delta_{2g}(\vec{p}) f_{g}(1+f_{g})\nonumber\\
\zeta_q&=&\frac{\nu_q}{3 T^2} \int \frac{d^3\vec{p}}{8\pi^3} \frac{p^2}{\omega_q^2} (p^2-3 c^2_s \omega_q^2)\Delta_{2q}(\vec{p}) f_{q}(1-f_{q})\nonumber\\
\zeta_s&=&\frac{\nu_s}{3 T^2} \int \frac{d^3\vec{p}}{8\pi^3} \frac{p^2}{\omega_s^2} (p^2-3 c^2_s \omega_s^2)\Delta_{2s}(\vec{p}) f_{s}(1-f_{s}).
\end{eqnarray}
Notice that while obtaining the expression for the bulk viscosity, we have 
exploited the Landau-Lifshitz (LL) Condition for the stress energy tensor. 
The factor $(-3 c^2_s \omega^2)$ in the right-hand side of Eq.(\ref{eq10}) is coming only because of that.
The appearance of this factor is not so straightforward. To obtain that one has to look for a particular solution of 
transport equation for $\zeta$ so that the viscous stress tensor satisfies LL condition.
Such a solution is  obtained by invoking the conservation laws, and thermodynamic relations in quite general way in ~\cite{kapu},
and valid in the present case at the level of formalism. The  modifications will appear only
in terms on new equilibrium distribution functions, and the 
modified dispersion relations, $\omega$. There 
is no such issue with the $\eta$ since physically it is associated with the response with the change in the 
shape of the system at constant volume, on the other hand $\zeta$ is linked with the volume expansion at a fixed shape.
Here, $c_s^2$ is the speed of sound square
extracted from the lattice data on $(2+1)$-flavor lattice QCD.
The determination of $\Delta_{1g,1q,1s}(p)$, and $\eta_{g,q,s}$ can easily be done following~\cite{chandra_eta1,chandra_eta2}, and  $\Delta_{2g,2q,2s}(\vec{p})$ and $\zeta_{g,q,s}$ following~\cite{chandra_bulk}.

\subsection{Determination of the perturbative, $\Delta_1$ and $\Delta_2$}
To obtain a analytic expression for the perturbations, $\Delta_{1,2}$, in our 
analysis, one need to first set up the transport equation 
in the presence of turbulent color fields. This has been 
done in~\cite{bmuller,chandra_eta1,chandra_eta2} in the recent past. 
Here, we only quote the linearized transport equation, with Vlasov-Dupree diffusive term, which arise 
after considering the ensemble average over turbulent color fields, in the light cone frame.
 The transport equation thus obtained reads, 
\begin{equation}
\label{eq11}
v^\mu \frac{\partial}{\partial x^\mu} f_{eq}(p)+ {\bf V}_A f_1 f_{eq}(p)(1\pm f_{eq}(p))=0,
\end{equation}

where ($f_{eq} \equiv f_g,f_q,f_s$), and $v^\mu\equiv (1, \vec{v}_p)$, where $\vec{v}_p=\partial_{\vec p} \omega$ is the quasi-particle velocity. 
It is easy to realize that the quasi-particle model does not change the group velocity of the quasi-partons. Note that Eq. (\ref{eq11}) 
is written in the absence of collision term, and assuming the weak coupling approximation. 

The mathematical structure of the Vlasov-Dupree operator
is as follows,
\begin{equation}
\label{eq12}
{\bf V}_A=\frac{g^2 C_2}{2(N_c^2-1) \omega^2}<E^2+B^2> \tau_m {\bf L^2},
\end{equation}  
where $C_2$ is the quadratic Casimir invariant for partons. For gluons, $C_2=N_c$, and for quarks $C_2=(N_c^2-1)/{2 N_c}$.
Here, $\omega\equiv \lbrace \omega_g,\omega_q,\omega_s \rbrace$, denotes the quasi-partons dispersions,
and $g^2$ is the QCD coupling constant at finite temperature.  
The quantities $E^{a}$ and $B^{a}$ denotes the chromo field strengths, where {\it a} is the SU(3) color index, and
$<E^2+B^2> \equiv <E^a\cdot E^a+B^a\cdot B^a>$. The bracket $<..>$ denotes the ensemble average over the color field configurations
which are turbulent (grow in time with a time scale $\tau_m$) as described in ~\cite{bmuller}. The anomalous transport coefficients 
in this approach are obtained by invoking the argument that soft color fields are turbulent. 
Their action on quasi-partons can be described by considering the ensemble average over the color fields that leads to an effective 
Force term in the linearized transport equation.
The parameter, $\tau_m$ is the time scale associated with instability in the field,  and the operator ${\bf L^{2}}$ is,
\begin{eqnarray}
\label{eq13}
{\bf L^2}&=&-(\vec{p}\times\partial_{\vec p})^2+(\vec{p}\times\partial_{\vec p})\vert_z^2\nonumber\\
      &&\equiv -(L^{p})^2+({L^{p}}_{z})^2.\nonumber\\
\end{eqnarray}
Since ${\bf L^2}$ contains angular momentum operator $L^{p}$, therefore it gives non-vanishing 
contribution while operating on an anisotropic function of $\vec{p}$. It will always lead to the vanishing contribution 
while operating on a isotropic function of $\vec{p}$.
Following ~\cite{chandra_eta2}, the expression for the $\Delta_{1g}(p)$ is obtained as, 
\begin{equation}
\label{eq14}
\Delta_{1g}(p)=\frac{2(N_c^2-1) \omega_g^2\ T}{3 C_g g^2 <E^2+B^2>\tau_m}.
\end{equation}

On the other hand, expressions for $\Delta_{1q,1s}$
are obtained as,
\begin{eqnarray}
\label{eq15}
\Delta_{1q}(p)&=&\frac{2(N_c^2-1) \omega_q^2\ T}{3 C_f g^2 <E^2+B^2>\tau_m}\nonumber\\
\Delta_{1s}(p)&=&\frac{2(N_c^2-1) \omega_s^2\ T}{3 C_f g^2 <E^2+B^2>\tau_m}.
\end{eqnarray}

Now, we write the transport equation containing only those terms which contribute to 
bulk viscosity $\zeta$ as,
\begin{eqnarray}
\label{eq16}
(\frac{p^2}{3 \omega^2}&-&c^2_s)\frac{\omega}{T}(\nabla\cdot\vec{u})f_{eq}(1\pm f_{eq})
\nonumber\\&=&\frac{g^2 C_2}{3(N_c^2-1) \omega^2}<E^2+B^2> \tau_m {\bf L}^2 \ f_1(\vec{p},\vec{r})f_{eq}(1\pm f_{eq}).\nonumber\\
\end{eqnarray}
Following \cite{chandra_bulk}, we can obtain the mathematical forms of 
the corresponding perturbations, $\Delta_2$. We shall write down the expressions in the
gluonic sector, and matter sector separately to avoid the confusion. 
The expression for $\Delta_{2g}(p)$ is obtained as, 
\begin{equation}
\label{eq17}
\Delta_{2g}(\vec{p})=\frac{4(N_c^2-1)T
\omega_g^2}{N_c g^2 <E^2+B^2> \tau_m p^2 }(\frac{p^2}{3}-c^2_s\ \omega_g^2) \ln(\frac{p_T}{\sqrt{6} T}) 
\end{equation}

On the other hand, the expressions for $\Delta_{2q,2s}$ are obtained as,
\begin{eqnarray}
\label{eq18}
\Delta_{2q}(\vec{p})=\frac{4(N_c^2-1)T \omega_q^2}{C_2 g^2 <E^2+B^2> \tau_m p^2 }(\frac{p^2}{3}-c^2_s \omega_q^2) \ln(\frac{p_T}{\sqrt{6} T})\nonumber\\ 
\Delta_{2s}(\vec{p})=\frac{4(N_c^2-1)T \omega_s^2}{C_2 g^2 <E^2+B^2> \tau_m p^2 }(\frac{p^2}{3}-c^2_s\ \omega_s^2) \ln(\frac{p_T}{\sqrt{6} T}).\nonumber\\ 
\end{eqnarray}

Next, we relate the denominator of Eqs.(\ref{eq15}), (\ref{eq17}), and (\ref{eq18}) to the 
parton energy loss parameter $\hat{q}\equiv \hat{q}_g,\ \hat{q}_q$, via the relation~\cite{majumdar},
\begin{equation}
\label{eq19}
\hat{q}=\frac{2 g^2 C_2}{3(N_c^2-1)} <E^2+B^2> \tau_m.
\end{equation}

The relation of $\hat{q}$, with the transport parameters in the present analysis is 
attributed to the fact that radiative energy loss  ($\hat{q}$ being a measure) depends on the 
rate of momentum exchange between the fast parton and the QCD medium. More precisely,
$\hat{q}$ is assumed as a rate of growth of the transverse momentum fluctuations of a fast parton 
to an ensemble of turbulent color fields, expressed as in Eq. (\ref{eq19}).

Now the gluonic,  contributions to $\eta$, and $\zeta$
in terms of $\hat{q}$ can be rewritten as follows,
\begin{eqnarray}
\label{eq20}
\eta_g&=&  \frac{T^6}{\hat{q}} \frac{64(N_c^2-1)}{ 3\pi^2} PolyLog[6,z_g], \nonumber\\
\zeta_g &=& \frac{4(N_c^2-1)}{3 T \pi^2 \hat{q}} \int\int p_{T} dp_{T} dp_{z} (\frac{p^2}{3}-c^2_s\omega_g^2)^2\times\nonumber\\
 && \ln(\frac{p_T}{p_0})\times f_{g}(1+f_{g}).
\end{eqnarray}
On the other hand, quark-antiquark viscosities in the matter sector are 
obtained as,

\begin{eqnarray}
\label{eq21}
\eta_{q}&=& \frac{64 N_c^2 \nu_{q}}{3 \pi^2 \hat{q} (N_c^2-1)} \lbrace -PolyLog[6,-z_q]\rbrace\nonumber\\
\eta_{s}&=& \frac{64 N_c^2 \nu_{s}}{3 \pi^2 \hat{q} (N_c^2-1)}\bigg \lbrace-PolyLog[6,-z_q]\nonumber\\&&+\frac{\tilde{m}^2}{2} PolyLog[5,-z_q]\bigg\rbrace\nonumber\\
\zeta_{q,s}&=& \frac{N_c\nu_{q,s} }{3 C_f T \pi^2 \hat{q}} \int \int  p_{T} dp_{T} dp_{z} (\frac{p^2}{3}-c^2_s \omega_{q,s}^2)^2\times\nonumber\\ && \ln(\frac{p_T}{p_0})\times f_{q,s}(1-f_{q,s}).
\end{eqnarray}
Here $\tilde{m}\equiv  m/T$ (mass of the strange quark scaled with temperature), and
the {\it PolyLog} functions that appear in the expressions for $\eta_{g,q,s}$ are defined  in terms of the series representation 
as,
\begin{equation}
\label{eq22}
Ploylog[n,x]=\sum_{k=1}^\infty \frac{x^k}{k^n},
\end{equation}
where n is a positive integer, and the convergence of the series is ensured by the fact that 
$x\leq 1$. Moreover, $PolyLog[n,1]\equiv \zeta(n)$, and also  $PolyLog[n,-1]\sim -\zeta(n)$.

Clearly from Eqs. (\ref{eq20}) and (\ref{eq21}), the various components of $\eta$, and $\zeta$ 
have strong dependence on the hot QCD EOS through the 
parameters $z_{g,q}$, and their first order derivatives with respect to temperatures, the speed of sound $c_s^2$, and 
$\hat{q}$ (speed of sound dependence is only there in $\zeta$). 
Therefore, before discussing the results for a
particular lattice EOS utilized in this analysis, it is instructive to discuss the dependence of lattice EOS 
on $\eta$ and $\zeta$ in view of the uncertainties in the height, and width of the interaction measure (trace anomaly)
computed in lattice QCD at finite temperature by different collaborations. The temperature dependence of $z_g$, and $z_q$
is mainly dependent on the temperature dependence of the interaction measure. The former, is directly related to the 
contributions coming from the gluonic action, and later depends on the interaction measure in (2+1)-flavor QCD subtracting 
gluonic contribution. Therefore, they both carry effects of lattice artifacts and uncertainties from the beginning of 
their determination. The same is true for $c_s^2$, since it has strong dependence upon the behavior of the interaction 
measure as a function of temperature. In fact, $c_s^2$ is related to the temperature derivative of the trace anomaly scaled
with the energy density~\cite{gupta_cs}. Therefore, it would be appropriate  to compare the 
predictions on $\zeta$, and $\eta$ based on the lattice data from various groups on the hot QCD EOS. 
However, this is beyond the scope of the present work, since we need lattice data from various lattice groups
not only for the full (2+1)-flavor QCD but also the contributions from the gluonic action to the EOS
within the same lattice computational setup, which is not an easy task to do. 
Moreover, it is not possible to use the pure SU(3) EOS since it shows 
at first order transition, in contrast to crossover shown by (2+1)-flavor QCD at vanishing baryon density.
Leaving aside the above comparison for future, we here only concentrate on a particular set of lattice data~\cite{cheng,dattas}. 
Since the magnitude, and the temperature behavior of $z_{g,q}$, and $c_s^2$ will change things mainly quantitatively, 
leaving intact some of the interesting physical observations (modulation of the $\eta$ as compare to the ideal EOS), and 
rapid decrease of $\zeta$ with increasing temperatures. Present analysis led us to strongly believe that there will a be strong impact of temperature dependence of 
interaction measure specifically on $\zeta$ and the ratio $\zeta/\eta$ for the temperatures closer to $T_c$.

 Next, the components of $\eta$ employing the ideal EOS for quarks and gluons (equivalently ideal
form of the their thermal distribution functions, which are nothing but the equilibrium distribution functions obtained by putting 
$z_{g,q}\equiv 1$ in Eq. (\ref{eq1})) can straightforwardly be obtained from Eqs. (\ref{eq20}) and (\ref{eq21}),
  by substituting $z_g\equiv 1$ and $z_q\equiv 1$.
To denote these components, the superscript {\it Id} (stands for the ideal EOS) is used. We thus obtain,

\begin{eqnarray}
\label{eq21s}
\eta_g^{Id}&=&  \frac{T^6}{\hat{q}} \frac{64(N_c^2-1)}{ 3\pi^2} \zeta(6),\nonumber\\
\eta_{q}^{Id}&=& \frac{T^6}{\hat{q}} \frac{64 N_c^2 \nu_{q}}{3 \pi^2 (N_c^2-1)}\times \frac{31}{32} \zeta(6) \rbrace,\nonumber\\
\eta_{s}^{Id}&=& \frac{T^6}{\hat{q}} \frac{64 N_c^2 \nu_{s}}{3 \pi^2 (N_c^2-1)}\bigg \lbrace\frac{31}{32}\zeta(6)\nonumber\\&&+\frac{\tilde{m}^2}{2}\times \frac{15}{16}
\zeta(5) \bigg\rbrace.
\end{eqnarray}
Here, following relations have been utilized $-PolyLog[5,-1]=\frac{15}{16} \zeta(5)$, and  $PolyLog[6,1]\equiv \zeta(6)\equiv -\frac{32}{31} PolyLog[6,-1]$. To appreciate the above expressions more, we can redo the whole analysis with $z_{g,q}\equiv 1$ and  unmodified dispersion relations 
$\omega_{g,q}=p$; $\omega_s=\sqrt{p^2+m^2}$, we shall end up with the ideal components of $\eta$ displayed
in Eq. (\ref{eq21s}). The expressions in Eq. (\ref{eq21s}) will be utilized in 
the next section while investigating the role  of interactions.

\subsection{Landau-Lifshitz condition and the bulk viscosity}
Here, we shall briefly describe the LL condition to obtain the 
form of the expression for $\zeta$ given in Eq. (\ref{eq10}).
We shall argue below that the solution thus obtained follows the LL 
condition adopting a recent analysis of Charkobarty, and Kapusta~\cite{kapu}.
Inputs have also been taken from the recent work of Dusling, and Sch\"{a}fer~\cite{dush1}, and 
Dusling and Teaney~\cite{dush2} regarding the viscous hydrodynamics.

Recall that LL matching condition is a way to specify uniquely, 
$\epsilon$, and $u^\mu$ in terms of the components of $T^{\mu\nu}$.
In LL convention,
\ba
\epsilon=u^{\mu}u^{\nu} T^{\mu\nu}\nonumber\\
\epsilon u^\mu=u^\nu T^{\mu\nu}.
\ea
The other six independent component of $T^{\mu\nu}$ are obtained by a non-equilibrium 
viscous stress $\Pi^{\mu\nu}=\pi^{\mu\nu}-\Delta^{\mu\nu} \Pi$ that satisfy 
$u_\mu \Pi^{\mu\nu}=0$. It is sufficient that this condition is satisfied in the 
LRF. This can be translated in to the fact that energy-shift due to the 
non-equilibrium terms vanishes. Denoting this energy shift by $\delta \epsilon$, we obtain the 
following condition,
\be
\label{ess}
\delta \epsilon=0=\sum_a \int \frac{d^3\vec{p}}{8\pi^3} \omega \delta f,
\ee
where $a$ sums over $g$, $q$ and $s$ here. As stated earlier $\omega$ and $\delta f$ is the
combined notations for non-equilibrium part of the distribution function for
these three sectors. Here, we have considered the medium modified dispersion for the single particle 
energy to implement the interaction correctly. This is also in same spirit as 
in the case of effective mass quasi-particle models described in~\cite{dush1}.
Such effects are encoded in 
form of $\delta f $ through $\Delta_1$, and $\Delta_2$ is the present case.
This condition can straightforwardly be satisfied in the case of shear viscosity due to the 
specific form of the $\pi^{\mu\nu}$. The non-trivialities are there in the bulk viscosity sector,
that we discuss below.

Next, using Eqs. (\ref{eq16}-\ref{eq18}) we can write Eq. (\ref{ess}) in the presence of the bulk viscosity 
as,
\ba
\label{esh}
\delta \epsilon =\sum_a  \int \frac{d^3 \vec{p}}{8\pi^3} \omega^2 \bigg(\frac{p^2}{3}-c_s^2 \omega^2 \bigg)
\tilde{\Delta}_2 f_{eq} (1\pm f_{eq}). 
\ea
From the expression for $\Delta_2$ in Eqs. (\ref{eq17}, \ref{eq18}), one can easily 
read off $\tilde{\Delta}_2$ as,
\be
\tilde{\Delta}_2=\frac{4(N_c^2-1) \omega}{ T \tau C_{2}g^2 <E^2+B^2> \tau_m } \ln(\frac{p_T}{\sqrt{6} T}). 
\ee
Here, $C_2$ denotes the respective quadratic Casimir invariants of $SU(N_c)$.

The energy shift in Eq. (\ref{esh}) will vanish iff $\omega^2 \tilde {\Delta}_2$ happens to be independent of 
$\omega$, and $\vec{p}$~\cite{dush1} that is based on the definition of the speed of sound ($c_s^2=\frac{\partial P}{\partial \epsilon}$ at constant ${\mathcal S}$).  In this case, Eq. (\ref{esh}) will read,
\ba
\label{esha}
\sum_a \int \frac{d^3 \vec{p}}{8\pi^3} \bigg(p^2-3 c_s^2 \omega^2 \bigg)
f_{eq} (1\pm f_{eq})=0.
\ea
The above condition can not be achieved with the $\omega$ dependence of $\tilde{\Delta}_2$ in present case.
It will be useful while obtaining expression for $\zeta$, invoking the LL condition below.
In the case collisional processes only, the quantity $\tilde{\Delta}_2$ is 
closely related to the relaxation time which is obtained in term of inverse the transport cross-section~\cite{dush1}.
Clearly, our particular solution for $\zeta$ obtained by solving 
the effective transport equation does not satisfy the LL condition.

Next, we discuss how one gets a physically relevant solution based on this particular solution for 
the $\zeta$ that satisfies LL condition. To that end, we closely follow a recent analysis of~\cite{kapu}.
Let us now define a quantity  $A_a(\omega)$ for the computational  convenience here as,
\be
A_a(\omega)= \frac{\omega}{3} (p^2-3 c_s^2 \omega^2)\tilde{\Delta}_2. 
\ee

Recall that $\omega\equiv\lbrace \omega_g,\omega_q,\omega_s\rbrace$, and 
$f_{eq}\equiv\lbrace f_g,f_q,f_s\rbrace$.

In this notation bulk viscosity, $\zeta$ will have the following expression (in terms of the particular solution),
\be
\label{ztt}
\zeta =\frac{1}{3} \sum_a  \int \frac{d^3 \vec{p}}{8\pi^3 \omega } p^2 f_{eq}(1\pm f_{eq}) A_a(\omega).
\ee

Now following~\cite{kapu}, we can consider a shift in $A_a(\omega)$ as,
$A_a (\omega)\rightarrow A^\prime_a (\omega)=A_a (\omega)-b \omega$ in the absence of conserved charges, and 
chemical potentials. This generates other set of solutions with coefficient $b$ being arbitrary. This leads to
the following expression for $\zeta$,

\be
\label{ztta}
\zeta =\frac{1}{3} \sum_a  \int \frac{d^3 \vec{p}}{8\pi^3 \omega } p^2 f_{eq}(1\pm f_{eq}) 
(A_a(\omega)-b\omega).
\ee

Now to fix $b$, we demand that the new solution must satisfy LL condition. This translates in to 
the LL condition for the new solution using Eq. (\ref{esha}) as,
\be
\label{ll}
\sum_a \int \frac{d^3 \vec{p}}{8\pi^3} \omega  \bigg(A_a(\omega)-b \omega\bigg)f_{eq}(1\pm f_{eq})=0.
\ee

Now, recast Eq. (\ref{ll}) as,
\ba
\label{ll1}
&&\sum_a \int \frac{d^3 \vec{p}}{8\pi^3} 3 b  c_s^2 \omega^2 f_{eq}(1\pm f_{eq})\nonumber\\
&=&\sum_a \int \frac{d^3 \vec{p}}{8\pi^3} 3c_s^2 \omega  A_a(\omega)f_{eq}(1\pm f_{eq}).
\ea

Using the condition given in Eq. (\ref{esha}), we obtain,
\ba
\label{ll2}
&&\sum_a \int \frac{d^3 \vec{p}}{8\pi^3 \omega} b\omega p^2  f_{eq}(1\pm f_{eq})\nonumber\\
&=& \sum_a \int \frac{d^3 \vec{p}}{8\pi^3} 3c_s^2 \omega  A_a(\omega)f_{eq}(1\pm f_{eq}).
\ea

Substituting Eq. (\ref{ll2}) in to Eq. (\ref{ll}), we obtain the bulk viscosity, 
$\zeta$:
\be
\label{zll}
\zeta= \frac{1}{3} \sum_a \int \frac{d^3 \vec{p}}{8\pi^3 \omega} f_{eq}(1\pm f_{eq}) A_a(\omega)(p^2-3c_s^2 \omega^2). 
\ee
Now, writing $\zeta$ in the component forms in Eq. (\ref{zll}), we
eventually reached to the desired expressions for $\zeta$ which are quoted in Eq. (\ref{eq10}).
Let us now proceed to investigate the temperature dependence of $\eta$ and $\zeta$.

\section{Temperature dependence of $\eta$ and $\zeta$}
The determinations of $\eta$, and $\zeta$ in the 
gluonic and matter sector, are incomplete unless to 
fix the temperature dependence of $\hat{q}$ in both the sectors.
The determination of $\hat{q}$ has been presented in the 
various phenomenological studies~\cite{hatq}, either based on 
the eikonal approximation, or the higher twist approximation, at a particular
value of the temperature. Here, we choose the $\hat{q}$ for gluons as $4.5\ Gev^2/fm$, and $2.0 Gev^2/fm$
for quarks, at $T=0.4 Gev$~\cite{hatq1} (this temperature, we denote as $T_0$). 
Since, $\hat{q}$ appears in the denominator in the expressions for $\eta$ and $\zeta$. Therefore, any set of values higher then those 
mentioned above will further decrease the values of $\eta$ and $\zeta$. 
At $T=T_0$, we can see that $\hat{q}_{g}=2.25 \hat{q}_{q}$. At this juncture, we do not know these parameters at all temperatures, 
so we assume this relation holds for all temperatures. This assumption is based on the definition of $\hat{q}$ in the 
leading order in hot QCD~\cite{techqm}, where its same for both gluons and quarks except that of the quadratic Casimir factor.
We shall utilize the relation $\hat{q}_g=2.25 \hat{q}_q$, while studying the temperature dependence of 
various quantities in the next subsections
The exact temperature dependence of $\hat{q}$, employing 
the quasi-particle description of hot QCD is not known to us at the moment.
This will be a matter of future investigations.

\subsection{Relative contributions}
In the section, discussions are mainly on,
(i) relative contributions of various components of  $\eta$
with their ideal counter parts,  (ii) gluonic verses matter sector for 
$\eta$, and $\zeta$ respectively. 

Note that the  shear and bulk viscosities 
in the (2+1)-flavor can be obtained by summing of all the 
individual contributions of the quasi-partons as,
\begin{eqnarray}
\label{eq23}
\eta&=&\eta_g +\eta_q +\eta_s\nonumber\\
\zeta&=&\zeta_g+\zeta_q+\zeta_s.
\end{eqnarray}
The additivity of various components here is attributed 
to the fact that all of them belong to same process, {\it viz.} the anomalous transport.
Viscosity contributions from distinct processes (e.g. anomalous and collisional) are inverse
additive due to the fact that various rates~\cite{bmuller,chandra_bulk} are additive. 

Let us define the relative quantities of our interest.
Firstly, we shall define the ratios of various components of $\eta$ to that for the 
ideal system of quarks and gluons (denoted as $\eta^{Id}$, and displayed in Eq.(\ref{eq21s})), which are defined as follows,
\ba
\label{rel1}
R_{gi}&&\equiv\frac{\eta_g}{\eta_g^{Id}}; R_{qi,si}\equiv\frac{\eta_{q,s}}{\eta_{q,s}^{Id}}\nn
R_{i}&&\equiv\frac{(\eta_g+\eta_q+\eta_s)}{(\eta_g^{Id}+\eta_q^{Id}+\eta_s^{Id})}.
\ea

Similarly, to compare the relative contributions among various components of 
$\eta$, we define the following ratios,
\ba
\label{rel2}
R_{gq} \equiv \frac{\eta_g}{\eta_q};   R_{gs} \equiv \frac{\eta_g}{\eta_s}; R_{sq}&&\equiv \frac{\eta_s}{\eta_q}.
\ea

On the other hand, to compare the relative contributions among the various components of $\zeta$, following 
quantities have been defined,

\ba
\label{rel3}
 R^{gq}\equiv \frac{\zeta_g}{\zeta_q};  R^{gs}\equiv \frac{\zeta_g}{\zeta_s};  R^{sq}\equiv \frac{\zeta_s}{\zeta_q}. 
\ea

\begin{figure}
\vspace{2mm}
\includegraphics[scale=.40]{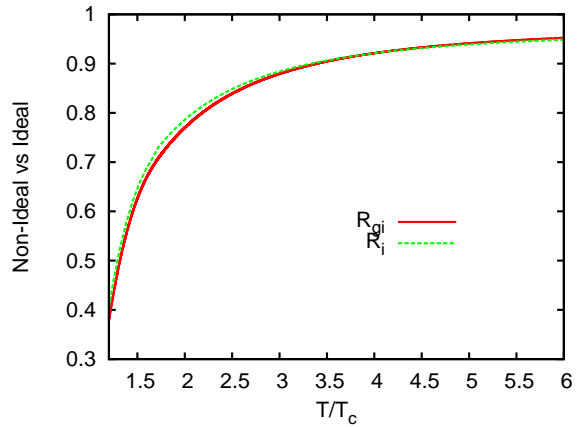} 
\caption{\label{fig1} (Color online) $\eta$ relative to the that obtained using the ideal EOS for QGP, in the gluonic sector, and
the  (2+1)-flavor is plotted as a function of $T/T_c$. The solid curve denotes the gluonic sector and dashed line denotes the 
(2+1)-flavor. Both $R_{gi}$ and $R_i$ approach to the ideal limit asymptotically.}
\vspace{2mm}
\end{figure}

\begin{figure}
\vspace{2mm}
\includegraphics[scale=.40]{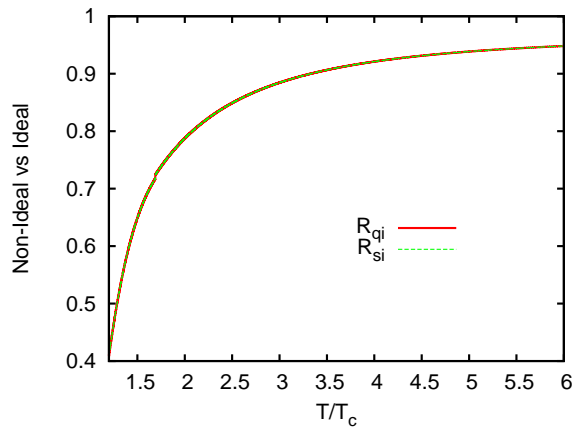} 
\caption{\label{fig2} (Color online) $\eta$ relative to the that obtained using the ideal EOS for the QGP, in the matter sector. The $R_{qi}$ is  $\eta$ relative to $\eta^{Id}$ in the light-quark sector, and similarly $R_{si}$ is for the strange quark sector.
Both the curves sits on the top of each other since the mass effects from the strange-quark sector do not play significant role here.} 
\vspace{2mm}
\end{figure}

\begin{figure}
\vspace{2mm}
\includegraphics[scale=.40]{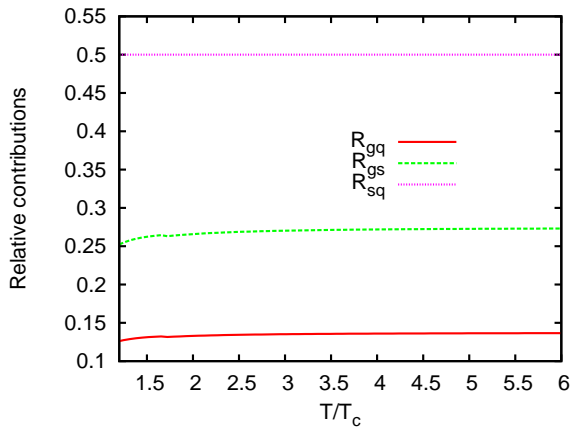} 
\caption{\label{fig3} (Color online) Shear viscosity in effective gluonic sector relative to matter sector.
The solid lines denotes $\eta_g$ relative to $\eta_q$, thin dashed lines (middle) represents $\eta_g$ relative to
$\eta_s$, and upper thick dashed line represents $\eta_s$ relative to $\eta_q$, as a function of $T/T_c$.} 
\vspace{2mm}
\end{figure}

\begin{figure}
\vspace{2mm}
\includegraphics[scale=.40]{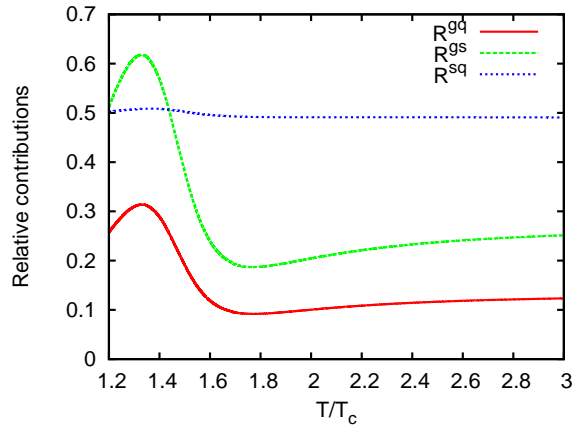} 
\caption{\label{fig4} (Color online) Temperature dependence of relative quasi-parton bulk viscosities.
The solid line  shows the behavior of $\zeta_g$ relative to $\zeta_q$,
thin dashed line shows the behavior of $\zeta_g$ relative to $\zeta_s$, and the 
thick dashed line shows the behavior of $\zeta_s$ relative to $\zeta_q$, as a function of 
$T/T_c$.} 
\vspace{2mm}
\end{figure}

The quantities defined in Eqs. (\ref{rel1}-\ref{rel3}) have been shown as a functions of $T/T_c$, in 
Figs. \ref{fig1}-\ref{fig4}.
The ratios $R_{gi}$ and $R_{i}$ are shown as a function of temperature in 
Fig. \ref{fig1}. The parameter $\hat{q}$ is assumed to be same in the interacting and ideal sector.
We have considered temperature dependence beyond $1.2 T_c$. Both $R_{gi}$, and $R_{i}$ show that interactions 
significantly modify the shear viscosity in the gluonic sector and the (2+1)-flavor QCD at lower temperatures.
Both of them lie within the range $\lbrace 0.40, 0.97\rbrace$ for the temperature range, 
$\lbrace T/T_c=1.2, 6.0\rbrace$. $R_{qi}$ and $R_{si}$ are shown in Fig. \ref{fig2} as a function of temperature.
Both of them sit on the top of each other. This is not surprising since the mass effects coming from the strange 
quark sector contribute negligibly in the temperature range considered here. The light quark sector and strange 
quarks differ with each other by a factor of two coming from the degrees of freedom. While considering the ratio
$R_{si}$ it cancels from the numerator and denominator. From Fig.\ref{fig2}, it is evident that the hot
 QCD interactions significantly modify the 
shear viscosity in the matter sector same as in the gluonic sector as compared to the ideal counter parts.
All of them approaches asymptotically to the ideal limit which is nothing but {\it unity}.
These observations suggest that $\eta$ could be thought of as a good diagnostic tool to 
distinguish various equations of state at the RHIC and the LHC.

Next, we investigate the gluonic shear and bulk viscosities relative to that of the 
matter sector. The relevant quantities in this context of $\eta$ are $R_{gq}$, $R_{gs}$, and $R_{sq}$ given in Eq.(\ref{rel2}).
These are shown as a function of temperature in Fig. \ref{fig3}. On the other hand for $\zeta$, $R^{qg}$, $R^{qs}$, and  $R^{sq}$
are shown as a function of temperature in Fig. \ref{fig4}. It can be observed from Fig. \ref{fig3}, and Fig. \ref{fig4} that 
the matter sector contributions significantly dominate over the gluonic contributions as far as the $\eta$ and $\zeta$ are concerned.
This could perhaps be understood by the following facts, {\it viz.}, the higher transport rates in the gluonic sector as compared to quark sector as encoded in $\hat{q}$, and the interactions entering through the effective fugacities $z_g$ and $z_q$.
Quantitatively, $\eta_g$ is $\sim 0.125 \eta_q$, and  $0.250 \eta_s$ at $T=1.20 T_c$, and increases quite slowly as a function 
of $T/T_c$ reaching around $0.135\ \eta_q$ around $6 T_c$ (see Fig. \ref{fig3}). The $\eta_s$ almost stays $0.5\ \eta_q$ for the considered range of temperature (contribution from the strange quark mass is almost negligible). From Fig. \ref{fig4}, 
it can be observed that $R^{gq}$ and $R^{gs}$ have same qualitative behavior as a function of temperature.
The quantitative difference is because of a factor $\sim 2$, since $\zeta_s\sim 0.5 \zeta_q$. Again the 
mass effects in the strange quark-sector play almost negligible role. The ratio $R^{gq}$ initially increases and attains a peak around 
$T/T_c \sim 1.37$ and then decreases sharply until $T/T_c=1.6$ and slightly increases beyond $1.6$ and indicating towards the saturation at higher temperatures. Quantitatively, $\zeta_g \approx 0.27 \zeta_q$ around $1.2 T_c$, and $ 0.13 \zeta_q$ at $3.0 T_c$.
These observations are very crucial in deciding the temperature dependence of $\eta$ and $\zeta$, and the ratios $\eta/{\mathcal S}$, $\zeta/{\mathcal S}$ and $\zeta/\eta$. Most of the recent studies devoted to the $\eta$ and $\zeta$ draw inferences for the QGP which are purely based on the study of the pure 
$SU(3)$ sector of QCD only. The matter sector has largely been ignored. In the light of the above observations, it is not 
desirable to exclude the matter sector since the dominant contributions are from there. 
Finally, we can obtain the exact value of the ratios $\eta/{\mathcal S}$, and $\zeta/{\mathcal S}$ by employing the 
values of $\hat{q}$ quoted earlier  ($\hat{q}=4.5 GeV^2/fm$ for gluons and $2.0 GeV^2/fm$ for quarks at $T=400 MeV$).
The ratio, $\eta/{\mathcal  S}$ thus obtained as $0.570$ and $\zeta/{\mathcal S}$ came out to be $0.057$ at $T=400 MeV$.
As discussed earlier, to obtain the exact temperature dependence of $\eta$, and $\zeta$, 
one requires to fix the temperature dependence of $\hat{q}$ within the quasi-particle model employed here.
This will be taken up separately in the near future. The quantity which can be determined unambiguity is the 
ratio $\zeta/\eta$ which is very crucial in deciding when the hot QCD becomes conformal. In other words, until what value of the 
temperature the effects coming from $\zeta$ are important while studying the QGP. 
We shall now proceed to discuss these issues next.

\subsection{The ratio $\zeta/\eta$}
\begin{figure}
\vspace{2mm}
\includegraphics[scale=.40]{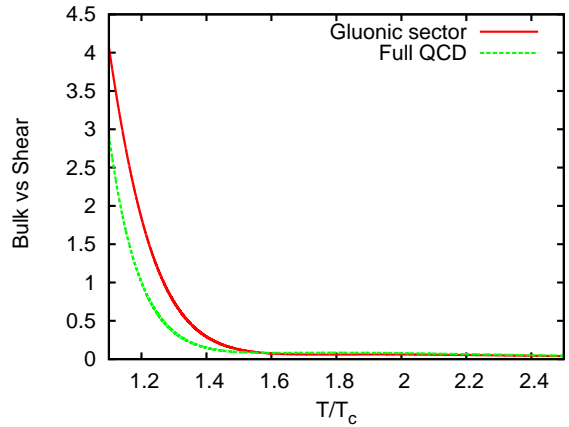} 
\caption{\label{fig5} (Color online) Temperature dependence of the ratio $\zeta/\eta$, in the effective gluonic sector, and the (2+1)-flavor QCD.
The solid line represents the gluonic sector, and the dashed line represents the (2+1)-flavor case.}
\vspace{2mm}
\end{figure}

\begin{figure}
\vspace{2mm}
\includegraphics[scale=.40]{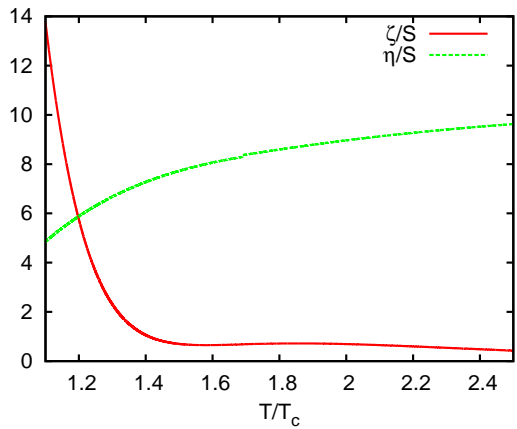} 
\caption{\label{fig6} (Color online) The temperature dependence of the quantities, $\eta/S$ and $\zeta/S$ in the (2+1)-flavor QCD. Here 
$S=\frac{{\mathcal S} T^3}{\hat{q}}$, where ${\mathcal S}$ denotes the entropy density.}
\vspace{2mm}
\end{figure}

The behavior of the 
ratio $\zeta/\eta$ as a function of temperature is shown in Fig. \ref{fig5}, and the temperature dependence of the ratios $\frac{\eta}{S}\equiv \frac{\eta \hat{q}}{T^3 {\mathcal S}}$ and $\frac{\zeta}{{\mathcal S}}\equiv \frac{\zeta\hat{q}}{T^3 {\mathcal S}}$, is shown in Fig. \ref{fig6}. 

Most importantly, from Fig. \ref{fig5},
clear indications are observed that $\zeta$ in the gluonic sector, and the (2+1)-flavor QCD diverge
as we approach closer to $T_c$ (the results are not shown around $T_c$, since such a quasi-particle picture may not be 
valid there.). The quantity $\zeta/\eta$ shows sharp decrease until one reaches up to $1.4 T_c$ in the gluonic sector and 
$1.6 T_c$ in the (2+1)-flavor QCD sector. Beyond that the decrease becomes slow and the ratio slowly approaches to zero. 
Such a behavior of $\zeta/\eta$ as a function of temperature could mainly be described in the
formal expressions in Eqs. (\ref{eq20}), and (\ref{eq21}), and decided only by the temperature dependence of 
$c_s^2$ but also by the energy-dispersion relations, $\omega_{g,q,s}$, and the temperature dependence of 
the effective fugacities, $z_{g,q}$. It is evident that there is no way to obtain 
a $(c_s^2-\frac{1}{3})^2$ factor out from the expression while performing the integration. However, such a scaling could be realized 
whenever $p<< T^2 \partial_T (ln(z_{g,q}))$, and $\omega_{g,q,s}$ happen to be independent of $z_g$ and $z_q$, and the thermal distribution of 
quai-partons show near ideal behavior. It may perhaps be realized at a very high temperature which are not relevant to study the QGP 
in the RHIC and the LHC. Therefore, $\zeta/\eta$ obtained here does not follow either a 
quadratic scaling  or a linear scaling with the conformal measure $(c_s^2-\frac{1}{3})$. The same conclusions were obtained in 
the case of pure gauge theory recently~\cite{chandra_bulk}. Note that for the scalar field theories, $\zeta/\eta=15(c_s^2-\frac{1}{3})^2$ (quadratic scaling)~\cite{scalar}, and  it has been found to be true for a photon gas coupled with the matter~\cite{weinberg}. The quadratic scaling is also 
valid in the case of perturbative QCD with a proportionality factor different from $15$~\cite{moore_1}. Furthermore, 
in the case of near conformal theories with gravity duals, $\zeta/\eta$ shows linear dependence on ($c_s^2-\frac{1}{3}$)~\cite{confo}. 

Finally, 
in Fig. \ref{fig6}, $\frac{\eta}{S}$ and $\frac{\zeta}{S}$ are plotted as a 
function of $T/T_c$ (here the quantity $S$ is related to the entropy density (${\mathcal S}$) as $S=\frac{{\mathcal S}\hat{q}}{T^3}$.
For the entropy density, we utilize the quasi-particle results which are shown to be consistent with the 
predictions of lattice QCD, and in all the plots $c_s^2$ has been obtained from the quasi-particle model employing the method 
quoted in~\cite{gupta_cs}.  Interestingly, these are of same order at $T=1.2 T_c$. Below that temperature the latter dominates over the former 
and vice versa for $T\ge 1.2 T_c$. The former increases, in contrast to latter as a function of $T/T_c$. There is a sharp increase shown by the latter until one reaches $1.4 T_c$, and beyond that the decrease is slower and one is quite close to the conformal limit of QCD.
The important inference that could be drawn from here is that while studying the QGP one needs to incorporate the effects of both shear and bulk viscosities until approximately $1.5 T_c$. This confirms our viewpoint that both $\eta$ and $\zeta$ have a significant impact on the properties of the QGP at the RHIC and the LHC.

\section{Conclusions and future prospects}
In conclusion, the shear and bulk viscosities of the hot QCD are estimated 
by combining a semi-classical transport equation with a quasi-particle realization of the 
(2+1)-flavor lattice QCD. The effective gluonic sector contributes an order of magnitude lower as 
compared to the matter sector while determining the transport coefficients of the hot QCD and the QGP.
This could perhaps be understood in terms of transport cross-sections of gluons and quark-antiquarks.
Since transport coefficients are inversely proportional to the cross-sections. The bulk viscosity of the (2+1)-flavor QCD
is found to be equally significant as the shear viscosity while modeling the QGP.
Indications are seen regarding a blow up in the bulk viscosity as we
go closer to $T_c$. 

The temperature dependence of the ratio $\zeta/\eta$ suggests that the QGP becomes almost conformal around $1.4 T_c-1.5 T_c$. The ratio sharply decreases from $T=1.1 T_c - 1.4 T_c$, and 
beyond that slowly approaches to zero. Therefore, in this regime we can ignore the effects of $\zeta$ while studying the 
hydrodynamic evolution and properties of the QGP. We further found that $\eta$ and $\zeta$ are of same order around $T=1.2 T_c$.
For temperatures lower than that $\zeta$ is dominant and for higher temperatures, $\eta$ is dominant.
Importantly, both $\eta$ and $\zeta$ came out to be highly sensitive to the presence of interactions.
This can be visualized from the modulation of $\eta$, as compared to its ideal counter part, and 
large and rising value of $\zeta$ for the temperatures that are closer to $T_c$ (due to 
large interaction measure there). The above conclusions are based on the fact that the
ratio $\hat{q}_g/\hat{q}_q$ is temperature independent which is approximately true 
with the definition of $\hat{q}$ considered in the present analysis (leading order in perturbative QCD). 
A generalization of the definition of the $\hat{q}$ in view of the quasi-particle picture may induce both 
qualitative and quantitative modifications to the ratio $\zeta/\eta$, and will be investigated in the near future.
 
It would be a matter of immediate future investigation to utilize
the more recent lattice data,
and compare the predictions for the  data from the hot QCD collaboration~\cite{leos_lat}, and the
Wuppertal-Budapest collaboration~\cite{leos1_lat}.
This would indeed be helpful in understating the impact of lattice artifacts, and uncertainties
on the transport properties of the QGP.

The investigations on the other contributions to the shear and bulk viscosities (collisional {\it etc.}), 
and their interplay with the corresponding anomalous transport coefficients will be a matter of future investigations. 
It will be interesting to include the effects of non-vanishing baryon density to the transport coefficients of 
the QGP. Moreover, one could include the anomalous transport coefficients in the Boltzmann-transport theory approach and 
study the impact on the response functions and quarkonia physics along the 
lines of ~\cite{chandra_nucla,chandra_iitr}, as well as dilepton production at the 
RHIC and the LHC. These ideas will be studied in the near future.

\vspace{2mm}
\section*{Acknowledgements} 
VC is highly thankful to R. S. Bhalerao and V. Ravishankar for encouragement and helpful discussions, 
F. Karsch and S. Datta for providing the lattice QCD data in the past. 
He sincerely acknowledges M. Maggiore, and the Department of Theoretical Physics, 
University of Geneva Switzerland, for the hospitality, where a significant part of this work was completed.
He is thankful to Dr. B. N. Tiwari for reading the manuscript and helping in the language part. 
He would like to acknowledge the financial support of the Tata Institute of Fundamental Research Mumbai, 
India in terms of a Visiting Fellow position, and INFN, Italy for awarding an INFN postdoctoral fellowship. 
He would further like to acknowledge the hospitality and financial support of the CERN-Theory division, Switzerland through the visitor program, and he is indebted to the people of India for their invaluable support for the research in basic
sciences in the country.

\end{document}